\documentstyle[12pt]{article}
\newcommand\bea{\begin{eqnarray}}
\newcommand\eea{\end{eqnarray}}

\begin{document}
\thispagestyle{empty}
\bibliographystyle{unsrt}
\setlength{\baselineskip}{18pt}
\parindent 24pt

\begin{center}{
{QUANTUM DECOHERENCE OF THE DAMPED HARMONIC OSCILLATOR } \vskip
0.5truecm
A. Isar \\
{Institute of Physics and Nuclear Engineering,
Bucharest-Magurele, Romania }\\
e-mail: isar@theory.nipne.ro
}
\end{center}

\begin{abstract}
In the framework of the Lindblad theory for open quantum systems, we
determine the degree of quantum decoherence of a harmonic oscillator
interacting with a thermal bath. It is found that the system
manifests a quantum decoherence which is more and more significant
in time. We also calculate the decoherence time and show that it has
the same scale as the time after which thermal fluctuations become
comparable with quantum fluctuations.
\end{abstract}


\section{Introduction}

The quantum to classical transition and classicality of quantum
systems continue to be among the most interesting problems in many
fields of physics, for both conceptual and experimental reasons
\cite{1,2}. Two conditions are essential for the classicality of a
quantum system \cite{3}: a) quantum decoherence (QD), that means the
irreversible, uncontrollable and persistent formation of a quantum
correlation (entanglement) of the system with its environment
\cite{4}, expressed by the damping of the coherences present in the
quantum state of the system, when the off-diagonal elements of the
density matrix decay below a certain level, so that this density
matrix becomes approximately diagonal and b) classical correlations,
expressed by the fact that the Wigner function of the quantum system
has a peak which follows the classical equations of motion in phase
space with a good degree of approximation, that is the quantum state
becomes peaked along a classical trajectory. Classicality is an
emergent property of open quantum systems, since both main features
of this process -- QD and classical correlations -- strongly depend
on the interaction between the system and its external environment
\cite{1,2}.

The role of QD became relevant in many interesting physical
problems. In many cases one is interested in understanding QD
because one wants to prevent decoherence from damaging quantum
states and to protect the information stored in quantum states from
the degrading effect of the interaction with the environment.
Decoherence is also responsible for washing out the quantum
interference effects which are desirable to be seen as signals in
experiments. QD has a negative influence on many areas relying upon
quantum coherence effects, in particular it is a major problem in
quantum optics and physics of quantum information and computation
\cite{5}.

In this work we study QD of a harmonic oscillator interacting with
an environment, in particular with a thermal bath, in the framework
of the Lindblad theory for open quantum systems. We determine the
degree of QD and then we apply the criterion of QD. We consider
different regimes of the temperature of environment and it is found
that the system manifests a QD which in general increases with time
and temperature.

The organizing of the paper is as follows. In Sec. 2 we review the
Lindblad master equation for the damped harmonic oscillator and
solve the master equation in coordinate representation. Then in Sec.
3 we investigate QD and in Sec. 4 we calculate the decoherence time
of the system. We show that this time has the same scale as the time
after which thermal fluctuations become comparable with quantum
fluctuations. A summary and concluding remarks are given in Sec. 5.

\section{Master equation and density matrix}

In the Lindblad axiomatic formalism based on quantum dynamical
semigroups, the irreversible time evolution of an open system is
described by the following general quantum Markovian master equation
for the density operator $\rho(t)$ \cite{6}:
\begin{eqnarray}{d \rho(t)\over dt}=-{i\over\hbar}[H,\rho(t)]
+{1\over 2\hbar} \sum_{j}([  V_{j} \rho(t), V_{j}^\dagger ]+[ V_{j},
\rho(t) V_{j}^\dagger ]).\label{lineq}\end{eqnarray} The harmonic
oscillator Hamiltonian $H$ is chosen of the general quadratic form
\begin{eqnarray} H=H_{0}+{\mu\over 2}(qp+pq), ~~~  H_{0}={1\over
2m}p^2+{m\omega^2\over 2}  q^2 \label{ham} \end{eqnarray} and the
operators $V_{j},$ $ V_{j}^\dagger,$ which model the environment,
are taken as linear polynomials in coordinate $q$ and momentum $p.$
Then the master equation (\ref{lineq}) takes the following form
\cite{7}:
\begin{eqnarray} {d \rho \over dt}=-{i\over \hbar}[ H_{0}, \rho]-
{i\over 2\hbar}(\lambda +\mu) [q, \rho p+ p \rho]+{i\over
2\hbar}(\lambda -\mu)[  p,
\rho   q+  q \rho]  \nonumber\\
-{D_{pp}\over {\hbar}^2}[  q,[  q, \rho]]-{D_{qq}\over {\hbar}^2} [
p,[  p, \rho]]+{D_{pq}\over {\hbar}^2}([  q,[  p, \rho]]+ [ p,[ q,
\rho]]). ~~~~\label{mast}   \end{eqnarray} The diffusion
coefficients $D_{pp},D_{qq},$ $D_{pq}$ and the dissipation constant
$\lambda$ satisfy the fundamental constraints: $ D_{pp}>0, D_{qq}>0$
and $D_{pp}D_{qq}-D_{pq}^2\ge {\lambda}^2{\hbar}^2/4.$ In the
particular case when the asymptotic state is a Gibbs state $
\rho_G(\infty)=e^{-{  H_0\over kT}}/ {\rm Tr}e^{-{ H_0\over kT}}, $
these coefficients become
\begin{eqnarray} D_{pp}={\lambda+\mu\over 2}\hbar
m\omega\coth{\hbar\omega\over 2kT}, ~~D_{qq}={\lambda-\mu\over
2}{\hbar\over m\omega}\coth{\hbar\omega\over 2kT}, ~~D_{pq}=0,
\label{coegib}
\end{eqnarray} where $T$ is the temperature of the thermal bath. In
this case, the fundamental constraints are satisfied only if
$\lambda>\mu$ and
\begin{eqnarray} (\lambda^2-\mu^2)\coth^2{\hbar\omega\over 2kT}
\ge\lambda^2\label{cons}\end{eqnarray} and the asymptotic values
$\sigma_{qq}(\infty),$ $\sigma_{pp}(\infty),$ $\sigma_{pq}(\infty)$
of the dispersion (variance), respectively correlation (covariance),
of the coordinate and momentum, reduce to \cite{7}
\begin{eqnarray} \sigma_{qq}(\infty)={\hbar\over
2m\omega}\coth{\hbar\omega\over 2kT}, ~~\sigma_{pp}(\infty)={\hbar
m\omega\over 2}\coth{\hbar\omega\over 2kT}, ~~\sigma_{pq}(\infty)=0.
\label{varinf} \end{eqnarray}

We consider a harmonic oscillator with an initial Gaussian wave
function ($\sigma_q(0)$ and $\sigma_p(0)$ are the initial averaged
position and momentum of the wave packet) \begin{eqnarray}
\Psi(q)=({1\over 2\pi\sigma_{qq}(0)})^{1\over 4}\exp[-{1\over
4\sigma_{qq}(0)}
(1-{2i\over\hbar}\sigma_{pq}(0))(q-\sigma_q(0))^2+{i\over
\hbar}\sigma_p(0)q], \label{ccs}\end{eqnarray} representing a
correlated coherent state (squeezed coherent state) with the
variances and covariance of coordinate and momentum
\begin{eqnarray} \sigma_{qq}(0)={\hbar\delta\over 2m\omega},~~
\sigma_{pp}(0)={\hbar m\omega\over 2\delta(1-r^2)},~~
\sigma_{pq}(0)={\hbar r\over 2\sqrt{1-r^2}}.
\label{inw}\end{eqnarray} Here $\delta$ is the squeezing parameter
which measures the spread in the initial Gaussian packet and $r,$
with $|r|<1$ is the correlation coefficient. The initial values
(\ref{inw}) correspond to a minimum uncertainty state, since they
fulfil the generalized uncertainty relation $
\sigma_{qq}(0)\sigma_{pp}(0)-\sigma_{pq}^2(0) =\hbar^2/4.$ For
$\delta=1$ and $r=0$ the correlated coherent state becomes a Glauber
coherent state.

From Eq. (\ref{mast}) we derive the evolution equation in coordinate
representation: \begin{eqnarray} {\partial\rho\over\partial
t}={i\hbar\over 2m}({\partial^2\over\partial q^2}-
{\partial^2\over\partial q'^2})\rho-{im\omega^2\over
2\hbar}(q^2-q'^2)\rho\nonumber\\
-{1\over 2}(\lambda+\mu)(q-q')({\partial\over\partial
q}-{\partial\over\partial q'})\rho+{1\over
2}(\lambda-\mu)[(q+q')({\partial\over\partial
q}+{\partial\over\partial
q'})+2]\rho  \nonumber\\
-{D_{pp}\over\hbar^2}(q-q')^2\rho+D_{qq}({\partial\over\partial
q}+{\partial\over \partial q'})^2\rho -{2iD_{pq}\hbar}(q-q')(
{\partial\over\partial q}+{\partial\over\partial
q'})\rho.\label{cooreq}\end{eqnarray} The first two terms on the
right-hand side of this equation generate the usual Liouvillian
unitary evolution. The third and forth terms are the dissipative
terms and have a damping effect (exchange of energy with
environment). The last three are noise (diffusive) terms and produce
fluctuation effects in the evolution of the system. $D_{pp}$
promotes diffusion in momentum and generates decoherence in
coordinate $q$ -- it reduces the off-diagonal terms, responsible for
correlations between spatially separated pieces of the wave packet.
Similarly $D_{qq}$ promotes diffusion in coordinate and generates
decoherence in momentum $p.$ The $D_{pq}$ term is the so-called
"anomalous diffusion" term and it does not generate decoherence.

The density matrix solution of Eq. (\ref{cooreq}) has the general
Gaussian form \begin{eqnarray} <q|\rho(t)|q'>=({1\over
2\pi\sigma_{qq}(t)})^{1\over 2} \exp[-{1\over
2\sigma_{qq}(t)}({q+q'\over
2}-\sigma_q(t))^2\nonumber\\
-{\sigma(t)\over 2\hbar^2\sigma_{qq}(t)}(q-q')^2
+{i\sigma_{pq}(t)\over \hbar\sigma_{qq}(t)}({q+q'\over
2}-\sigma_q(t))(q-q')+{i\over
\hbar}\sigma_p(t)(q-q')],\label{densol} \end{eqnarray} where
$\sigma(t)\equiv\sigma_{qq}(t)\sigma_{pp}(t)-\sigma_{pq}^2(t)$ is
the Schr\"odinger generalized uncertainty function. In the case of a
thermal bath we obtain the following steady state solution for
$t\to\infty$ ($\epsilon\equiv{\hbar\omega/2kT}$):
\begin{eqnarray} <q|\rho(\infty)|q'>=({m\omega\over
\pi\hbar\coth\epsilon})^{1\over 2}\exp\{-{m\omega\over
4\hbar}[{(q+q')^2\over\coth\epsilon}+
(q-q')^2\coth\epsilon]\}.\label{dinf}\end{eqnarray}

\section{Quantum decoherence}

An isolated system has an unitary evolution and the coherence of the
state is not lost -- pure states evolve in time only to pure states.
The QD phenomenon, that is the loss of coherence or the destruction
of off-diagonal elements representing coherences between quantum
states in the density matrix, can be achieved by introducing an
interaction between the system and environment: an initial pure
state with a density matrix which contains nonzero off-diagonal
terms can non-unitarily evolve into a final mixed state with a
diagonal density matrix.

Using new variables $\Sigma=(q+q')/2$ and $\Delta=q-q',$ the density
matrix (\ref{densol}) becomes \begin{eqnarray}
\rho(\Sigma,\Delta,t)=\sqrt{\alpha\over \pi}\exp[-\alpha\Sigma^2
-\gamma\Delta^2
+i\beta\Sigma\Delta+2\alpha\sigma_q(t)\Sigma+i({\sigma_p(t)\over\hbar}-
\beta\sigma_q(t))\Delta-\alpha\sigma_q^2(t)],\label{ccd3}\end{eqnarray}
with the abbreviations \begin{eqnarray} \alpha={1\over
2\sigma_{qq}(t)},~~\gamma={\sigma(t)\over 2\hbar^2
\sigma_{qq}(t)},~~ \beta={\sigma_{pq}(t)\over\hbar\sigma_{qq}(t)}.
\label{ccd4}\end{eqnarray}

The representation-independent measure of the degree of QD \cite{3}
is given by the ratio of the dispersion $1/\sqrt{2\gamma}$ of the
off-diagonal element $\rho(0,\Delta,t)$ to the dispersion
$\sqrt{2/\alpha}$ of the diagonal element $\rho(\Sigma,0,t):$
\begin{eqnarray} \delta_{QD}(t)={1\over 2}\sqrt{\alpha\over
\gamma}={\hbar\over 2\sqrt{\sigma(t)}}.\label{qdec}\end{eqnarray}

The finite temperature Schr\"odinger generalized uncertainty
function has the expression \cite{8} (with the notation
$\Omega^2\equiv\omega^2-\mu^2$,
$\omega>\mu$)\begin{eqnarray}\sigma(t)={\hbar^2\over
4}\{e^{-4\lambda
t}[1-(\delta+{1\over\delta(1-r^2)})\coth\epsilon+\coth^2\epsilon]\nonumber\\
+e^{-2\lambda t}\coth\epsilon[(\delta+{1\over\delta(1-r^2)}
-2\coth\epsilon){\omega^2-\mu^2\cos(2\Omega
t)\over\Omega^2}\nonumber\\ +(\delta-{1\over\delta(1-r^2)}){\mu
\sin(2\Omega t)\over\Omega}+{2r\mu\omega(1-\cos(2\Omega
t))\over\Omega^2\sqrt{1-r^2}}]+\coth^2\epsilon\}.\label{sunc}\end{eqnarray}
In the limit of long times Eq. (\ref{sunc}) yields
$\sigma(\infty)=(\hbar^2\coth^2\epsilon)/4,$ so that we obtain
\begin{eqnarray} \delta_{QD}(\infty)=\tanh{\hbar\omega\over
2kT},\end{eqnarray} which for high $T$ becomes
$\delta_{QD}(\infty)=\hbar\omega/2kT.$ We see that $\delta_{QD}$
decreases, and therefore QD increases, with time and temperature,
i.e. the density matrix becomes more and more diagonal at higher $T$
and the contributions of the off-diagonal elements get smaller and
smaller. At the same time the degree of purity decreases and the
degree of mixedness increases with $T.$ For $T=0$ the asymptotic
(final) state is pure and $\delta_{QD}$ reaches its initial maximum
value 1. $\delta_{QD}= 0$ when the quantum coherence is completely
lost, and if $\delta_{QD}= 1$ there is no QD. Only if
$\delta_{QD}<1$ we can say that the considered system interacting
with the thermal bath manifests QD, when the magnitude of the
elements of the density matrix in the position basis are peaked
preferentially along the diagonal $q=q'.$ Dissipation promotes
quantum coherences, whereas fluctuation (diffusion) reduces
coherences and promotes QD. The balance of dissipation and
fluctuation determines the final equilibrium value of $\delta_{QD}.$
The initial pure state evolves approximately following the classical
trajectory in phase space and becomes a quantum mixed state during
the irreversible process of QD.

\section{Decoherence time}

In order to obtain the expression of the decoherence time, we
consider the coefficient $\gamma$ (\ref{ccd4}), which measures the
contribution of non-diagonal elements in the density matrix
(\ref{ccd3}). For short times ($\lambda t\ll 1, \Omega t\ll 1$), we
have: \bea \gamma(t)=-{m\omega\over
4\hbar\delta}\{1+2[\lambda(\delta+{r^2\over\delta(1-r^2)})\coth\epsilon
+\mu(\delta-{r^2\over\delta(1-r^2)})\coth\epsilon-\lambda-\mu-{\omega
r\over\delta\sqrt{1-r^2}}]t\}.\label{td}\eea The quantum coherences
in the density matrix decay exponentially and the decoherence time
scale is given by \bea t_{deco}={1\over
2[\lambda(\delta+{r^2\over\delta(1-r^2)})\coth\epsilon
+\mu(\delta-{r^2\over\delta(1-r^2)})\coth\epsilon-\lambda-\mu-{\omega
r\over\delta\sqrt{1-r^2}}]}.\label{tdeco1}\eea The decoherence time
depends on the temperature $T$ and the coupling $\lambda$
(dissipation coefficient) between the system and environment, the
squeezing parameter $\delta$ and the initial correlation coefficient
$r.$ We notice that the decoherence time is decreasing with
increasing dissipation, temperature and squeezing.

For $r=0$ we obtain:\bea t_{deco}={1\over
2(\lambda+\mu)(\delta\coth\epsilon-1)}\label{tdeco2}\eea and at
temperature $T=0$ (then we have to take $\mu=0$), this becomes \bea
t_{deco}={1\over 2\lambda(\delta-1)}.\eea We see that when the
initial state is the usual coherent state $(\delta=1),$ then the
decoherence time tends to infinity. This corresponds to the fact
that for $T=0$ and $\delta=1$ the coefficient $\gamma$ is constant
in time, so that the decoherence process does not occur in this
case.

At high temperature, expression (\ref{tdeco1}) becomes ($\tau\equiv
{2kT/\hbar\omega}$) \bea t_{deco}={1\over
2[\lambda(\delta+{r^2\over\delta(1-r^2)})
+\mu(\delta-{r^2\over\delta(1-r^2)})]\tau}.\eea If, in addition
$r=0,$ then we obtain \bea t_{deco}={\hbar\omega\over
4(\lambda+\mu)\delta kT}.\eea

The generalized uncertainty function $\sigma(t)$ (\ref{sunc}) has
the following behaviour for short times: \bea
\sigma(t)={\hbar^2\over 4}\{1+ 2[\lambda
(\delta+{1\over\delta(1-r^2)})\coth\epsilon+\mu(\delta-{1\over\delta(1-r^2)})
\coth\epsilon-2\lambda]t\}.\label{sunc1}\eea This expression shows
explicitly the contribution for small time of uncertainty that is
intrinsic to quantum mechanics, expressed through the Heisenberg
uncertainty principle and uncertainty due to the coupling to the
thermal environment. From Eq. (\ref{sunc1}) we can determine the
time $t_d$ when thermal fluctuations become comparable with quantum
fluctuations. At high temperature we obtain \bea t_d={1\over
2\tau[\lambda
(\delta+{1\over\delta(1-r^2)})+\mu(\delta-{1\over\delta(1-r^2)})]}.\eea
By thermal fluctuations we mean the fluctuations that arise in the
generalized uncertainty function $\sigma(t)$ from the coupling of
the harmonic oscillator to the thermal bath at arbitrary temperature
$T,$ even at $T=0.$ By quantum fluctuations we mean fluctuations of
the quantum harmonic oscillator at zero coupling with the thermal
bath.

As expected, we see that the decoherence time $t_{deco}$ has the
same scale as the time $t_d$ after which thermal fluctuations become
comparable with quantum fluctuations \cite{9,10}. The values of
$t_{deco}$ and $t_d$ become closer with increasing temperature and
squeezing.

When $t\gg t_{rel},$ where $t_{rel}\approx\lambda^{-1}$ is the
relaxation time, which governs the rate of energy dissipation, the
particle reaches equilibrium with the environment. In the
macroscopic domain QD occurs very much faster than relaxation, so
that for all macroscopic bodies the dissipation term becomes
important much later after the decoherence term has already
dominated and diminished the off-diagonal terms of the density
matrix. We remark also that $t_{deco}$ can be of the order of
$t_{rel}$ for sufficiently low temperatures and small wave packet
spread (small squeezing coefficient).

\section{Summary and concluding remarks}

We have studied QD with the Markovian equation of Lindblad for a
system consisting of an one-dimensional harmonic oscillator in
interaction with a thermal bath in the framework of the theory of
open quantum systems based on quantum dynamical semigroups.

(1) Using the criterion of QD for the considered model, we have
shown that QD in general increases with time and temperature. For
large temperatures, QD is strong and the degree of mixedness is
high, while for zero temperature the asymptotic final state is pure.
With increasing squeezing parameter and initial correlation, QD
becomes stronger, but the asymptotic value of the degree of QD does
not depend on the initial squeezing and correlation, it depends on
temperature only. QD is expressed by the loss of quantum coherences
in the case of a thermal bath at finite temperature.

(2) We determined the general expression of the decoherence time,
which shows that it is decreasing with increasing dissipation,
temperature and squeezing. We have also shown that the decoherence
time has the same scale as the time after which thermal fluctuations
become comparable with quantum fluctuations and the values of these
scales become closer with increasing temperature and squeezing.
After the decoherence time, the decohered system is not necessarily
in a classical regime. There exists a quantum statistical regime in
between. Only at a sufficiently high temperature the system can be
considered in a classical regime.

The Lindblad theory provides a self-consistent treatment of damping
as a general extension of quantum mechanics to open systems and
gives the possibility to extend the model of quantum Brownian
motion. The results obtained in the framework of this theory are a
useful basis for the description of the connection between
uncertainty, decoherence and correlations (entanglement) of open
quantum systems with their environment, in particular in the study
of Gaussian states of continuous variable systems used in quantum
information processing to quantify the similarity or
distinguishability of quantum states using distance measures, like
trace distance and quantum fidelity.

\section*{Acknowledgments}

The author acknowledges the financial support received within the
Project PN 06350101/2006.

\end{document}